# Wire active microrheology to differentiate viscoelastic liquids from soft solids

**F. Loosli, M. Najm, R. Chan, E. Oikonomou, A. Grados, M. Receveur and J.-F. Berret***

*Matière et Systèmes Complexes, UMR 7057 CNRS Université Denis Diderot Paris-VII, Bâtiment Condorcet, 10 rue Alice Domon et Léonie Duquet, 75205 Paris, France.*

**Abstract:** Viscoelastic liquids are characterized by a finite static viscosity and a zero yield stress, whereas soft solids have an infinite viscosity and a non-zero yield stress. The rheological nature of viscoelastic materials has long been a challenge, and it is still a matter of debate. Here, we provide for the first time the constitutive equations of linear viscoelasticity for magnetic wires in yield stress materials, together with experimental measurements using Magnetic Rotational Spectroscopy (MRS). With MRS, the wires are submitted to a rotational magnetic field as a function of frequency and the wire motion is monitored by time-lapse microscopy. The soft solids studied are gel-forming polysaccharide aqueous dispersions (gellan gum) at concentrations above the gelification point. It is found that soft solids exhibit a clear and distinctive signature compared to viscous and viscoelastic liquids. In particular, the wire average rotation velocity equals zero over a broad frequency range. We also show the MRS technique is quantitative. From the wire oscillation amplitudes, the equilibrium elastic modulus is retrieved and agrees with polymer dynamics theory.

## 1. Introduction

In rheology, viscoelastic solids are defined as materials that behave like solids under weak applied stresses and as liquids at higher stresses.[1,2] These materials are characterized by a critical yield stress value, $\sigma_Y$ that separates a regime of pure deformation from that of deformation and flow. The measure of yield stresses has long been a challenge in rheology, and it is still a matter of intense debate.[3-5] Well-known yield stress solids are polymer and colloidal gels, foams, emulsions and pastes, which are materials of interest in many research fields including chemistry, pharmaceutics, agriculture or environment applications.[3,5-12] Viscoelastic solids are also termed soft solids, and we will use alternatively both terminologies.

In practice, soft solids with low yield stress values, of the order of 1 Pa or less are not able to support their own weight, and as a result they appear as flowing materials when their container is shaken or overturned. In such cases, the identification of a yield stress behavior, but also the determination of $\sigma_Y$ itself are both delicate.[3-5,9,10,13] To solve this problem, controlled-stress rheometers with improved performance have been developed and over the years have become highly sensitive instruments. Current detection thresholds are nowadays around $10^{-7}$ N m for torques and $10^{-8}$ s$^{-1}$ for the deformation rates.[4] In this context, going to low shear stresses is interesting because it permits to differentiate between deformation and flow regimes and hence to get a measure of the yield stress. Recent studies have shown however that deformations and/or flows observed at extremely low stresses such as those mentioned above are not always steady, and that instabilities such as wall slip and shear banding may occur, leading to erroneous readout of the rheological outcomes.[3,5] In addition, shearing solutions over extended periods of time poses practical problems such as those related to the instrumental stability or to the solvent evaporation. An alternative approach to reduce applied stresses, and to access the low shear rate range consists in reducing the size of the shearing device. For cone-and-plate and Couette devices for instance, transmitted torques vary as the diameter to the third power.[1,2] Being able to reduce the size of a measuring tool from 1 cm to 10 μm would result in a decrease of the applied torque by a factor $10^9$. In the following, we show that the reduction of the torque by several orders of magnitude is achievable combining the use of magnetic micron-sized wires as shearing device and of an active microrheology set-up for their actuation.[14-24]

Here we address the issue of yield stress behavior using a magnetic wire-based microrheology technique, also called Magnetic Rotational Spectroscopy (MRS).[25-30] MRS has been developed in a first step to measure the viscosity of Newton liquids confined in small volumes, or of samples that cannot be processed by rheometry. The technique is based on the use of micron-sized wires submitted to a rotational magnetic field. MRS has benefited in recent years from significant advances in materials science, for instance from the synthesis of novel magnetic probes such as wires, rods, needle-like aggregates or helices,[31-43] and also from the development of magnetic micro-swimmers that can be maneuvered in fluidic environments in a controlled manner.[23,30,44,45] Magnetic Rotational Spectroscopy consists in monitoring the wire motion as a function of the actuating frequency: below a critical cut-off noted $\omega_C$ the wire rotation is synchronous with the field, while above this frequency it exhibits back-and-forth oscillations and it is asynchronous (i.e. not synced with the excitation). In terms of average rotation frequency, it has been found that the response over a broad frequency range appears as a resonance peak centered on $\omega_C$ and similar to that found in mechanical systems.[46,47] From the peak position and the use of the relationship $\omega_C \sim \eta^{-1}$, where $\eta$ denotes the static viscosity, this technique was thereafter described as a spectroscopic method for probing fluid viscosity.[25,26,48-51] Applications of the MRS technique to nanocomposite thin films and ceramic matrices for characterization, guiding and alignment were recently reviewed, showing its versatility in materials science.[30,52]

In Refs.[29,53], the MRS technique was extended to evaluate the mechanical response of more complex materials, and in particular of viscoelastic liquids. Active microrheology experiments conducted on Maxwell fluid models were shown to be in good agreement with theoretical predictions in terms of time and frequency dependences. In addition to the angular frequency $\omega_C$, a second key parameter, noted $\theta_0$ was introduced in the model and it was shown to vary inversely proportional to the elastic modulus.[53] The method was further tested to investigate the rheology of the intracellular medium of living mammalian cells. Results on murine NIH/3T3 fibroblasts and human cervical carcinoma HeLa cells have demonstrated that the cytoplasm can be appropriately described as a viscoelastic liquid, with finite viscosity and a broad relaxation time distribution.[28]

In the present paper, we further develop this concept by providing for the first time the constitutive equation of linear viscoelasticity of magnetic wires in a yield stress material, and by comparing model predictions with experimental measurements. The soft solids selected for this study were gel-forming polysaccharide aqueous dispersions (gellan gum) at concentrations below and above the gelification point. In the gel phase, the samples studied were characterized by elastic moduli between 10 and 300 Pa and by yield stresses between 0.1 and 10 Pa. The result that emerges from this research is that soft solids exhibit a clear and distinctive signature in Magnetic Rotational Spectroscopy. The average rotation velocity equals zero over a broad frequency range ($10^{-2} - 10^2$ rad s$^{-1}$), in agreement with constitutive modeling. In particular, no measurable critical frequency could be determined from the measurements. These findings allow defining a criterion to differentiate unambiguously viscoelastic liquids from soft solids.

# 2. Wire rotation in viscoelastic media

## 2.1 – Maxwell and Standard Linear Solid model predictions

In a rotating magnetic excitation $H$ at frequency $\omega$, a superparamagnetic wire is submitted to a magnetic torque of the form:[48,49,54]

$$\Gamma_m(H) = \frac{\chi^2}{2(2+\chi)} \mu_0 V H^2 \sin\, 2(\omega t - \theta(t)) \tag{1}$$

where $\chi$ is the material magnetic susceptibility, $\mu_0$ the vacuum permeability, $V = \pi D^2 L/4$ the volume of the wire (of length $L$ and diameter $D$). In the sinus argument, $\theta$ describes the wire orientation. Immersed in a viscoelastic material, the wire experiences a viscous and an elastic torques $\Gamma_v$ and $\Gamma_e$ respectively, that hinder its rotation. The torques read:[53,55]

$$\Gamma_v = \frac{\pi\,\eta L^3}{3g\left(\frac{L}{D}\right)}\frac{d\theta_v}{dt} \; ; \; \Gamma_e = \frac{\pi G L^3}{3g\left(\frac{L}{D}\right)}\theta_e \tag{2}$$

In Eq. 2, $\eta$ is the static viscosity, $G$ the elastic modulus and $g\left(\frac{L}{D}\right)$ a dimensionless function of the anisotropy ratio $p = L/D$. In this study, we assume $g(p) = \ln(p) - 0.662 + 0.917/p - 0.05/p^2$.[56] To describe viscoelastic fluids and solids, rheology uses Maxwell and Kelvin-Voigt constitutive models for the linear response. These models are described as a Hookean spring and a dashpot in series for the Maxwell model, and in parallel for the Kelvin-Voigt model (Figure 1 and Supplementary Information S1).

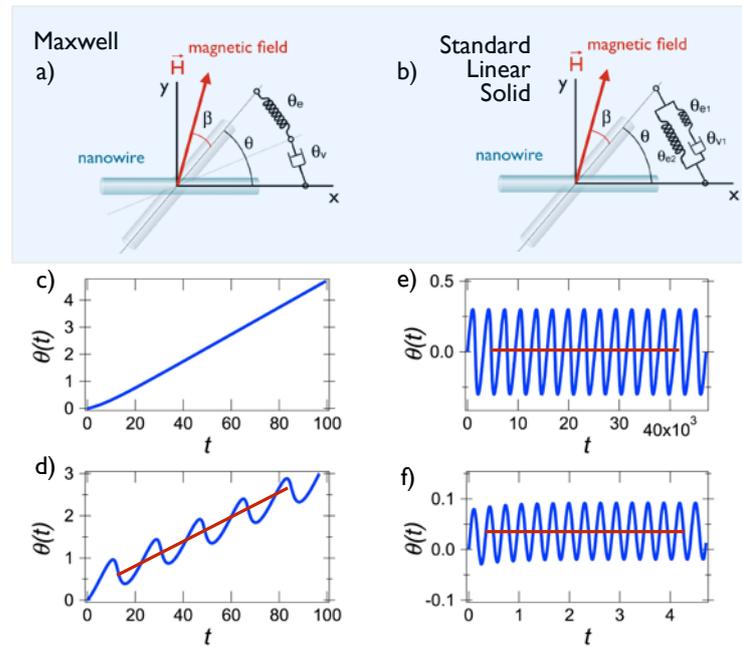

**Figure 1:** *a-b) Schematic representation of a wire actuated in a) a viscoelastic liquid and b) a viscoelastic solid. The Maxwell model is represented by a spring and a dashpot in series, whereas the Standard Linear Solid model is made of a Maxwell element and a spring in parallel. c-d) Time dependences of the wire orientation $\theta(t)$ calculated for the Maxwell model (Eq. 3) apart from the critical frequency $\omega_C$.[53] The figures illustrate the transition between the synchronous and asynchonous regimes observed as the actuating frequency is increased. The straight line in red denotes the average rotation frequency $\Omega$. e-f) Same as in s 1c-d for the Standard Linear Solid model at two different frequencies, $\omega = 10^{-3}$ and $10$ rad $s^{-1}$. Note that in this case $\Omega = 0$ (horizontal straight line in red).*

For the Maxwell model, the elastic and viscous deformations are additive, and the shear stresses coming from the separate elements are equal, leading to the equations $\theta = \theta_v + \theta_e$, and $\Gamma_m = \Gamma_v = \Gamma_e$. Using Eqs. 1 and 2, the wire rotation is described by the first order differential constitutive equation:[29,53]

$$\frac{d\theta(t)}{dt}(1 + \theta_0 \cos 2(\omega t - \theta)) = \omega_C \sin 2(\omega t - \theta) + \omega \theta_0 \cos 2(\omega t - \theta) \qquad (3)$$

where

$$\tau = \frac{\eta}{G}; \ \omega_C = \frac{3}{8}\frac{\mu_0 \Delta\chi}{\eta}\frac{H^2}{L^{*2}} \ ; \ \theta_0 = \frac{3}{4}\frac{\mu_0 \Delta\chi}{G}\frac{H^2}{L^{*2}} \qquad (4)$$

Here, $\tau$ denotes a relaxation time, $L^* = L/D\sqrt{g(L/D)}$, $\omega_C$ the critical frequency and $\theta_0$ the oscillation amplitude. Note that both $\omega_C$ and $\theta_0$ vary quadratically with $H/L^*$.

For the Kelvin-Voigt model, elastic and viscous torques are additive and oppose to the magnetic torque ($\Gamma_m = \Gamma_v + \Gamma_e$) and the angles are equal ($\theta = \theta_v = \theta_e$). The Kelvin-Voigt model is accurate in modeling creep experiments, but falls short to describe transients in controlled deformation experiments.[12] To circumvent this limitation, a modified Kelvin-Voigt model, dubbed Standard Linear Solid is generally preferred. The model combines a Maxwell element of viscosity $\eta$ and elasticity $G$ and a Hookean spring of elasticity $G_{eq}$ in parallel, as depicted in Figure 1b. Here $G_{eq}$ denotes the equilibrium storage modulus and characterizes the quasi-static (*i.e.* $\omega \to 0$) elastic response of a soft solid. The equation of motion for the wire is given by:

$$\frac{d\theta(t)}{dt}\left[1 + \frac{\theta_0}{\theta_{eq}} + \theta_1 cos2(\omega t - \theta)\right] + \frac{\theta_0}{\theta_{eq}\tau}\theta$$
$$= \theta_0\omega cos2(\omega t - \theta) + \frac{\theta_0}{2\tau}\sin 2(\omega t - \theta) \qquad (5)$$

where

$$\tau = \frac{\eta}{G}; \; \theta_0 = \frac{3}{4}\frac{\mu_0\Delta\chi}{G}\frac{H^2}{L^{*2}} \; \; ; \; \theta_{eq} = \frac{3}{4}\frac{\mu_0\Delta\chi}{G_{eq}}\frac{H^2}{L^{*2}} \qquad (6)$$

It can be verified that posing $G_{eq} = 0$ in the above equation yields the Maxwell equation in Eq. 3. The differential equations for the wire rotation in different model fluids were solved using the MatLab software (MathWorks).

## 2.2 – Wire rotation in model viscoelastic fluid and solid

Figures 1c–f compares the Maxwell and Standard Linear Solid constitutive models. We first examine the temporal motion of a wire submitted to a steady rotating magnetic field at frequency $\omega$. Figures 1c and 1d show the time dependence of the angle $\theta(t)$ for a Maxwell viscoelastic liquid at two frequencies apart from $\omega_C$. In the first low frequency regime, the wire rotation is synchronous with the field, and the wire orientation is delayed with respect to the field by the angle $\frac{1}{2}Arcsin(\omega/\omega_C)$.[48] For $\omega > \omega_C$ no stationary solution exists and the wire exhibits a back-and-forth motion as function of the time. The above predictions were tested on surfactant micellar solutions in the frequency range accessible to conventional rheometer, and excellent agreement between micro- and macrorheology was obtained.[29,53] Figures 1e and 1f display the rotation angle $\theta(t)$ for the Standard Linear Solid model at two representative frequencies, $\omega = 10^{-3}$ and 10 rad s$^{-1}$ respectively. Here, only one rotation regime is observed, namely that of periodic oscillations at frequency twice that of the excitation.

To simplify the analysis, the computed $\theta(t)$-traces are translated into a set of two parameters: the average rotation velocity $\Omega(\omega) = \langle d\theta(t)/dt\rangle_t$ and the back-and-forth oscillation amplitude $\theta_B(\omega) = \langle\theta_B(t, \omega)\rangle_t$. Figure 2a displays the average velocity $\Omega(\omega)$ *versus* frequency calculated for the Maxwell (Eq. 3) and Standard Linear Solid (Eq. 5) models. For the viscoelastic liquid, $\Omega(\omega)$ shows a cusp-like maximum centered at $\omega_C$. Its frequency behavior is given by the expression:

$$\omega \leq \omega_C \qquad \Omega(\omega) = \omega$$
$$\omega \geq \omega_C \qquad \Omega(\omega) = \omega - \sqrt{\omega^2 - \omega_C{}^2} \qquad (7)$$

The wire response function appears as a resonance peak similar to that found in mechanical systems[27,52] and the fluid viscosity is derived from the peak position using Eqs. 4 and 7. For the Standard Linear Solid model in contrast, it is found that whatever the rotation frequency:

$$\Omega(\omega) = 0 \qquad (8)$$

Viscoelastic liquids and yield stress gels exhibit hence very different responses as far as wire actuation is concerned. In this paper, we propose to use the wire average angular velocity as a criterion to differentiate viscoelastic liquids from soft solids.

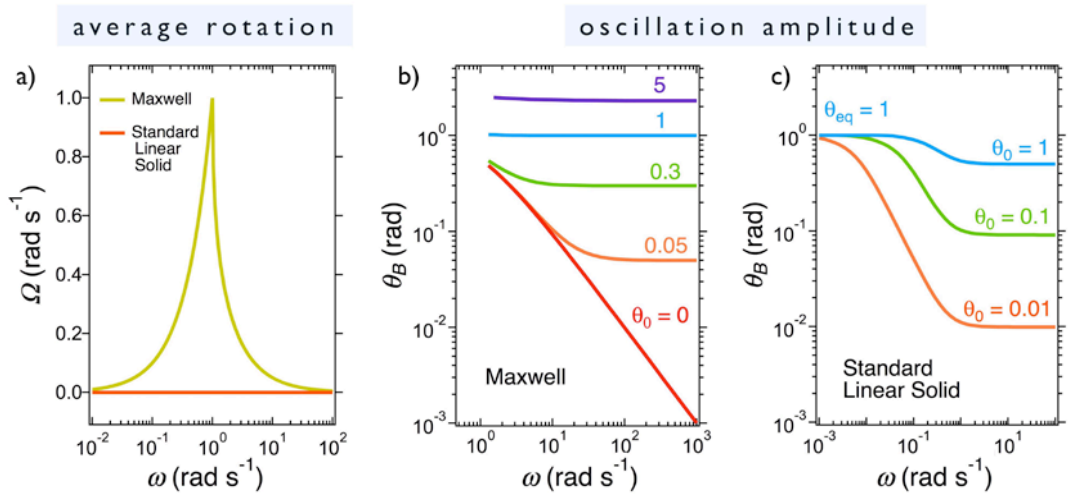

***Figure 2:*** *a) Average angular velocity $\Omega(\omega)$ as a function of the actuating frequency obtained for the Maxwell and for the Standard Linear Solid models. $\Omega(\omega) = \langle d\theta(t)/dt \rangle_t$ is computed from Eq. 3 and Eq. 5. For the viscoelastic liquid, a critical frequency $\omega_C$ is set at 1 rad s$^{-1}$. For the soft solid, $\theta_{eq} = 1$ rad and $\theta_0 = 0.1$ rad. b) Variation of angle $\theta_B(\omega)$ for the Newton ($\theta_0 = 0$) and for the Maxwell ($\theta_0 = 0.05, 0.3, 1$ and 5 rad) models, as computed from Eq. 3. c) Frequency dependence of the oscillation amplitudes $\theta_B(\omega)$ for the Standard Linear Solid model at $\theta_0 = 0.01, 0.1$ and 1 rad and $\theta_{eq} = 1$ rad (Eq. 5).*

Figures 2b and 2c illustrate the frequency dependence of the oscillation amplitudes $\theta_B(\omega)$ for the different models examined. In Figure 2b, $\theta_B(\omega)$ is calculated for the Maxwell fluid at different $\theta_0$-values (0, 0.05, 0.3, 1 and 5). Except for the case $\theta_0 = 0$ representing that of a purely viscous fluid, the angle decreases with increasing frequency and flattens into a frequency independent plateau. At high frequency, one has $\lim_{\omega \to \infty} \theta_B(\omega) = \theta_0$, where $\theta_0$ is given by Eq. 4. The high frequency plateau in the amplitudes is indeed the signature of the medium elasticity in this regime. Figure 2c displays calculated amplitudes for the Standard Linear Solid model at different $\theta_0 = 0.01, 0.1$ and 1, keeping $\theta_{eq} = 1$. In the examples considered, $\theta_B(\omega)$ exhibits a sigmoidal decrease between two asymptotic plateau regimes. For the Standard Linear Solid model, the low and high frequency limits are given by:

$$\lim_{\omega \to 0} \theta_B(\omega) = \theta_{eq} \ and \ \lim_{\omega \to \infty} \theta_B(\omega) = \frac{\theta_0 \theta_{eq}}{\theta_0 + \theta_{eq}} \qquad (9)$$

It is interesting to note the similarities between $\theta_B(\omega)$ and $1/G'(\omega)$ dependence: $1/G'(\omega)$ also displays a sigmoidal variation, decreasing from the inverse equilibrium elastic modulus $1/G_{eq}$ at low frequency to the high frequency $1/(G + G_{eq})$. In conclusion to this part, it is found that for Newton, Maxwell and Kelvin-Voigt models, $\Omega(\omega)$ and $\theta_B(\omega)$ display specific asymptotic behaviors *versus* frequency. For viscoelastic liquids, the average rotation velocity $\Omega(\omega)$ exhibits a marked maximum, whereas for a soft solid it is constant and equal to 0.

# 3. Results and discussion

### 3.1 – Polysaccharide gel rheology

Phytagel is a linear polysaccharide produced from *Pseudomonias elodea* bacterium.[57] With increasing concentration, the biopolymer solutions exhibit a sol-gel transition at a critical concentration $c$ = 0.2 wt. % (Figure 3a). In the gel, linear rheology experiments were performed as a function of the strain to determine the linear regime ($\gamma < 1\%$) and as a function of the frequency. As shown in Figures 3b and 3c, the elastic modulus is found to be higher than the loss modulus.

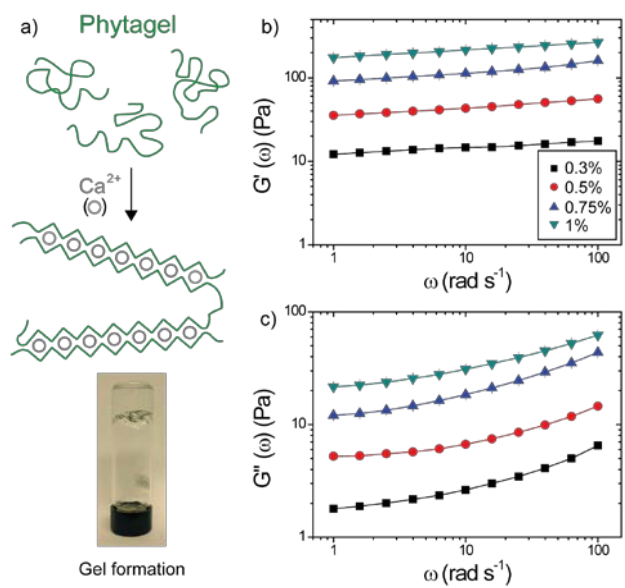

**Figure 3:** *a) Schematic illustration of the gel formation for linear polysaccharide chains in presence of divalent calcium ions. Ca²⁺ facilitate the gelation of phytagel solution due to its complexation with the carboxylate moieties. Image of a 1 wt. % phytagel solution at room temperature 30 min after the vial was turned over. b – c) represent the storage modulus $G'(\omega)$ and loss modulus $G''(\omega)$, respectively, for phytagel solutions (0.3 – 1 wt. %) obtained by conventional rheology with a 0.3 % deformation in frequency sweep mode.*

$G'(\omega)$ also displays scaling behaviors of the form $G'(\omega) \sim \omega^{0.10 \pm 0.02}$ over the whole frequency range.[58,59] Nonlinear cone-and-plate rheology was carried out to ascertain the yield stress $\sigma_Y$ in the gel phase. The shear stress *versus* shear rate curves measured between $10^{-3}$ and 100 s⁻¹ are shown in Figure S2. The $\sigma(\dot{\gamma})$-dependences were adjusted using the Herschel-Bulkley model: $\sigma(\dot{\gamma}) = \sigma_Y + K\dot{\gamma}^n$, where the $K$ is the consistency, and $n$ an exponent that allows for a viscosity to vary with the shear rate.[3,5] For the $c$ = 0.3, 0.5, 0.75 and 1 wt. % phytagel dispersions, yield stress values $\sigma_Y$ = 0.5, 0.8, 1.5 and 7.2 Pa were obtained respectively. These above results (scaling behavior for $G'(\omega)$ and existence of a yield at low shear rates) are strong indications of the gel-like character of the samples.[1]

## 3.2 – Wire rotation in the non-yield stress polysaccharide dispersion

To set a reference, we first investigated the wire behavior on a non-yield stress polysaccharide dispersion. To this aim, a calcium-free phytagel solution at concentration of 0.2 wt. % was prepared, resulting in a slightly viscoelastic liquid of viscosity 0.01 Pa s. Magnetic rotational spectroscopy was performed using a 90 μm long and 2.2 μm thick wire at temperature T = 26 °C at increasing frequency. In Figure 4a and 4b, the wire rotation angles $\theta(t)$ obtained at $\omega = 1.57$ and 1.76 rad $s^{-1}$ are displayed as a function of the time.

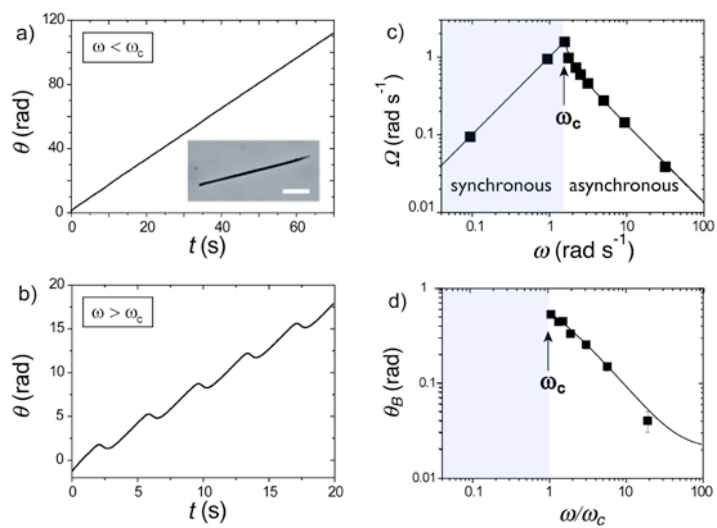

**Figure 4:** a – b) Wire orientation angle $\theta(t)$ in the synchronous ($\omega = 1.57$ rad $s^{-1}$) and asynchronous ($\omega = 1.76$ rad $s^{-1}$) regimes for a non cross-linked polysaccharide dispersion at c = 0.2 wt. %. Experiments were conducted at a magnetic field amplitude of 4 mT and a temperature of 26° C. Inset in 4a: optical microscopy image of a 90.0 μm wire used in the MRS experiment. c) Average angular velocity $\Omega(\omega)$ as a function of the frequency in a double logarithmic scale. The continuous line through the data points is obtained using Eq. 7. d) Oscillation amplitude $\theta_B(\omega)$ versus frequency in the asynchronous regime. The continuous line represents the Newton model predictions. From the average angular velocity and from the oscillation amplitude, the critical frequency was estimated $\omega_C = 1.7$ rad $s^{-1}$.

At low frequency, the wire rotates in phase with the field and the angle increases linearly with time. Above a critical frequency $\omega_C$, the wire shows a back-and-forth motion characteristic of the asynchronous regime, as described in the previous section. Figure 4c shows the average angular velocity $\Omega(\omega)$ in double logarithmic scale. With increasing frequency, $\Omega(\omega)$ passes through a maximum at $\omega_C = 1.67$ rad $s^{-1}$ before decreasing as $\Omega(\omega) \sim \omega^{-1}$. Least-square calculations using Eq. 7 provide an excellent fit to the data, and a static viscosity $\eta_0 = 0.012 \pm 0.01$ Pa s. This latter value is in good agreement with that of rheometry. Figure 4d displays the oscillation amplitude in the second rotation regime. There, $\theta_B(\omega)$ decreases with frequency in a similar fashion as for a viscoelastic liquid (using Eq. 3 with $\theta_0 = 0.02$ rad). Note that the continuous lines in Figure 4c and 4d were computed using $\omega_C$ as a single adjustable parameter.

## 3.3 - Time-resolved response in the gel phase

We now examine the wire motions in the polysaccharide gel phase. For each concentration studied between 0.3 and 1 wt. %, magnetic wires are dispersed in the gel and their motion is monitored by time-lapse microscopy under the application of a 9 mT rotating field (T = 30 °C). Depending on the applied frequency, 20 to 2000 s movies are recorded and digitalized, from which the center-of-mass and wire orientations are obtained and plotted *versus* time. Figure 5a display images of a 9.0 μm wire rotating in phytagel 0.5 wt. % at the angular frequency of 0.094 rad s⁻¹ (see entire movie in Supplementary).

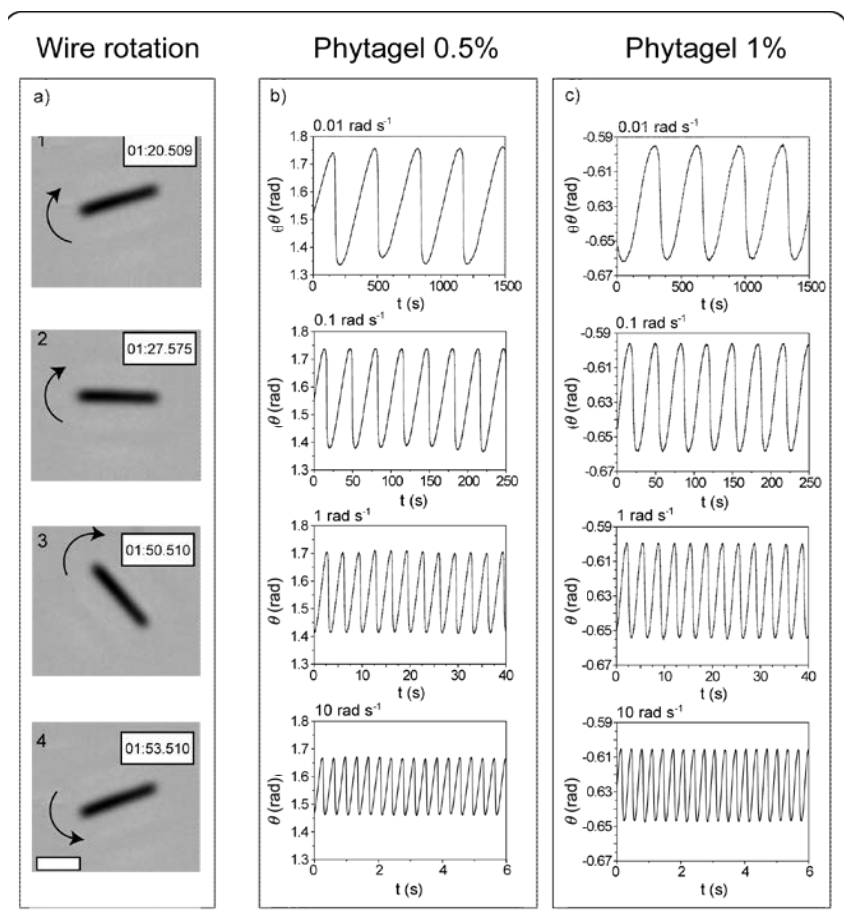

**Figure 5:** *a) Optical microscopy images of an actuated 9.0 μm wire in a 0.5 wt. % polysaccharide phytagel solution (frequency ω = 0.1 rad s⁻¹, scale bare 5 μm). Between images 3 and 4, the wire is subjected to a fast back motion, indicative of an asynchronous rotation regime. b – c) Rotation angle θ(t) for 18.9 μm wires dispersed in phytagel 0.5 wt. % and 1 wt. %, respectively. Note that the oscillation period is twice that of the actuating frequency (here 0.01, 0.1, 1 and 10 rad s⁻¹), in agreement with the constitutive models shown in Section II.*

During the first 30 seconds (image#1 to #3), the wire rotates at a constant rate in the clockwise direction (arrows), and then it comes back rapidly by 1.13 radian, indicating that the wire rotation is hindered. In the last image (image#4), the wire is back to its initial position, image#1 and #4 being superimposable. On a longer period, it is found that the wire is animated of oscillations characteristic of an asynchronous regime. This behavior is illustrated in Figures 5b and 5c for phytagel 0.5 and 1 wt. % respectively. There, the orientation angle θ(t) for 18.9 μm long wires is shown at various frequencies between $10^{-2}$ and 10 rad s⁻¹. In this range, a unique regime is observed, namely that of periodic oscillations characterized by a frequency twice that of the magnetic field. With increasing $\omega$, the oscillation amplitudes decreases slightly, from 0.40 to 0.20 rad for the 0.5 wt. % sample and from 0.07 to 0.04 rad for the 1 wt. % solution. As discussed in the next section, the oscillation amplitude reflects the strength of the gel elasticity, small amplitudes being related to large elastic moduli $G'(\omega)$.

### 3.4 – Average rotation frequency and oscillation amplitude in the gel phase

Figures 6a–d display the average angular velocity $\Omega(\omega)$ in the gel phase for wires comprised between 9 and 54 μm and for frequencies between $10^{-2}$ and $10^{2}$ rad s$^{-1}$ (T = 30° C). The wires characteristics are listed in Table I. Data show that for the different conditions examined the average velocity $\Omega(\omega)$ is comprised between $-2\times10^{-4}$ and $+7\times10^{-4}$ rad s$^{-1}$, *i.e.* much smaller that the actuating frequency. Zero average rotation speeds are in agreement with the Kelvin-Voigt and with the Standard Linear Solid models discussed in Section 2. The data in Figures 6a–d confirm hence the hypothesis of a gel-like rheology for calcium containing phytagel solutions. Results are also shown to be independent on the wire length in the studied range.

Figures 6e–h display the dependence of the oscillation amplitude $\theta_B(\omega)$ as a function of the frequency corresponding to the previous conditions. The major result here is that the angles $\theta_B(\omega)$ decreases with increasing wire length and gel concentration. For wires of approximately 10 μm, $\theta_B$ decreases from 1.3 to 0.1 rad between 0.3 and 1 wt. % (Table I). The oscillation amplitude variations for 10 μm wires are discussed in the next section. Similarly, at a fixed phytagel concentration (e.g. $c = 0.75$ wt. %, Figure 6g), the amplitude varies from 0.2 to 0.02 rad for wires increasing between 15 and 54 μm. These latter results are in good agreement with the asymptotic behaviors for the $\theta_0$ and $\theta_{eq}$-angles given in Eq. 6, which both decrease as $L^{-2}$. They also show that in the gel phase high moduli are associated with small amplitudes.

| $c$ | $L$ | $\Omega$ | $\theta_{eq}$ | $G_{eq}$ |
|:---:|:---:|:---:|:---:|:---:|
| wt. % | μm | rad s$^{-1}$ | rad | Pa |
| | 12.6 | $-1.9\times10^{-4}$ | 1.30 | 1.1 |
| 0.30 | 18.1 | $1.1\times10^{-4}$ | 1.01 | 2.5 |
| | 45.0 | $-5.1\times10^{-5}$ | 0.28 | 0.9 |
| | 9.0 | $7.2\times10^{-4}$ | 1.31 | 2.0 |
| 0.50 | 11.9 | $3.8\times10^{-5}$ | 0.56 | 3.5 |
| | 18.9* | $2.7\times10^{-5}$ | 0.42 | 2.7 |
| | 14.9 | $-1.6\times10^{-5}$ | 0.21 | 8.6 |
| 0.75 | 23.3 | $-1.2\times10^{-5}$ | 0.10 | 15 |
| | 54.9 | $3.7\times10^{-6}$ | 0.023 | 19 |
| | 10.7 | $5.3\times10^{-6}$ | 0.094 | 25 |
| 1.00 | 16.4 | $1.3\times10^{-5}$ | 0.085 | 17 |
| | 18.9* | $1.0\times10^{-5}$ | 0.066 | 19 |

**Table I**: *Experimental parameters obtained for gellan gum polysaccharide aqueous dispersions from magnetic rotational spectroscopy. c denote the polymer concentration, L the wire length, $\Omega$ the average rotation veolicity, $\theta_{eq}$ the oscillation amplitude extrapolated at zero frequency and $G_{eq}$ the equilibrium elastic modulus of the soft solids. Wires indicated with a star are related to the data shown in Figure 5 for c = 0.5 and 1 wt. %.*

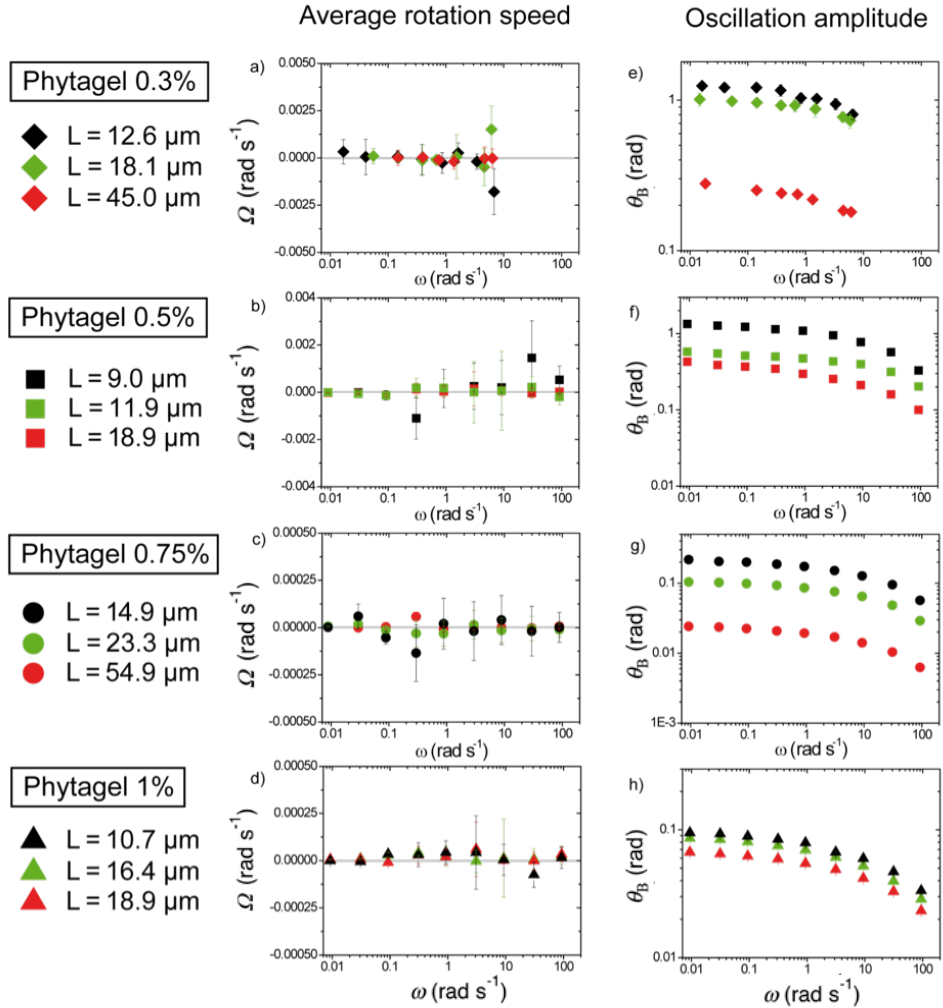

**Figure 6:** *a – d) Average rotation velocity $\Omega(\omega)$ of magnetic wires as a function of the frequency for phytagel concentrations $c = 0.3, 0.5, 0.75$ and $1$ wt. %, respectively. Horizontal straight lines in each panel indicate the behavior $\Omega(\omega) = 0$. e – f) Wire oscillation amplitude $\theta_B(\omega)$ versus frequency for the same phytagel concentrations. Black, green and red closed symbols represent wires of increasing lengths.*

## 3.5 – Equilibrium modulus determination

Predictions from the Standard Linear Solid model show that for asymptotic behaviors there exists a direct relationship between the modulus $G'(\omega)$ and the oscillation amplitude $\theta_B(\omega)$. At low frequency the equilibrium elastic modulus is given by $G_{eq} = \frac{3\mu_0\Delta\chi H^2}{4L^{*2}}\theta_B^{-1}(\omega \to 0)$, whereas at high frequency the instantaneous moduli reads $G = \frac{3\mu_0\Delta\chi H^2}{4L^{*2}}\left(\theta_B^{-1}(\omega \to \infty) - \theta_B^{-1}(\omega \to 0)\right)$. In the following, we use the Standard Linear Solid model to adjust the phytagel data and retrieve the elastic storage moduli and the relaxation times. Figure 7a displays the oscillation amplitude $\theta_B(\omega)$ for selected wires of length around 10 μm in the different gels studied. Least square calculations using model predictions are shown as continuous lines. Calculated $\theta_B(\omega)$ exhibit a plateau at low frequency, typically below $\omega = 10^{-1}$ rad s$^{-1}$ followed by a sigmoidal decrease towards a secondary plateau at high frequency, above $\omega = 10^2$ rad s$^{-1}$. Typical time constants obtained from the fits are of the order of 0.1 s. These times do not depend on the gel strength or on the concentration. At low frequency, a good agreement between experimental and calculated amplitudes is found. From the extrapolation at $\omega \rightarrow 0$, $\theta_{eq}$ is determined, leading to an estimation of the equilibrium modulus $G_{eq}$ (Eq. 6). $G_{eq}$-values are listed in Table I for the different conditions tested and displayed in Figure 7b as a function of the polysaccharide concentration. At higher frequency, the wire mechanical response is not accounted for by the model. The reason for this discrepancy is due to the fact that phytagel dispersions are characterized by a broad distribution of relaxation times (as suggested by the $G'(\omega)$ and $G''(\omega)$ behaviors), whereas the Standard Linear Solid model has a single relaxation time to describe viscoelasticity. In particular, the high frequency asymptotic behavior is not recovered, and the instantaneous elastic moduli $G$ cannot be evaluated. A systematic analysis made on different conditions leads to the data of Figure 7b. There, the concentration dependence of $G_{eq}$ and $G'$ obtained from with cone-and-plate at $\omega = 1$ rad s$^{-1}$ are compared. Both moduli exhibit scaling laws with an exponent 9/4, in agreement with polymer dynamics theory.[60] The prefactors before the $c^{9/4}$-scaling are 23 and 160 Pa, respectively, indicating that the polysaccharide gel equilibrium modulus is about 7 times lower than the storage modulus measured by cone-and-plate. The present findings show that the MRS technique is well suited to access rheological properties of gels not easily measurable using rheometry. Improved fitting using advanced soft solid models could be obtained by adding several Standard Linear Solid relaxators in parallel that would account for the relaxation time distribution of the material.

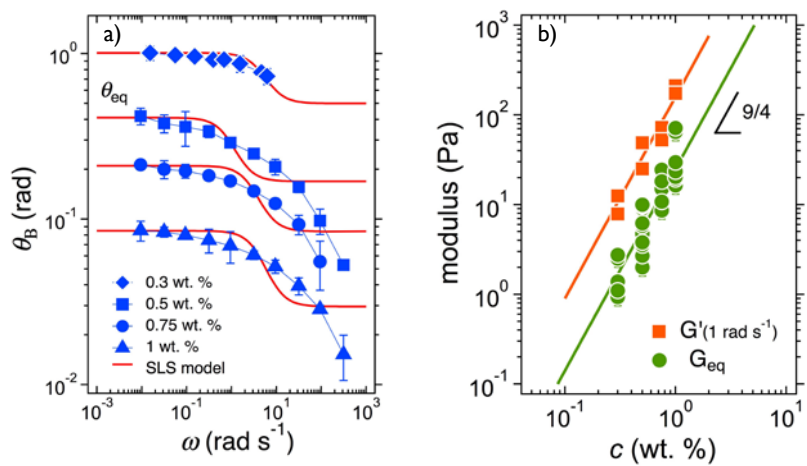

*Figure 7:* a) Oscillation amplitude $\theta_B(\omega)$ versus frequency for 10 μm wires dispersed in polysaccharide dispersion. Continuous lines in red are results from least-square calculations using the Standard Linear Solid model (Eq. 5). b) Storage and equilibrium modulus $G'(c)$ and $G_{eq}(c)$ respectively as a function of the polymer concentration. The $G'(c)$-values were obtained at $\omega = 1$ rad s$^{-1}$. Both moduli vary as $c^{9/4}$, as expected form polymer dynamics theory.

# 4. Conclusions

In this work, we use a novel microrheology technique named Magnetic Rotational Spectroscopy to determine the rheological nature of viscoelastic materials. Constitutive models of linear viscoelasticity used here are the Maxwell model and the Standard Linear Solid model. As shown theoretically in Section 2, the average rotation velocity of actuated wires exhibits distinctive behaviors for the two types of materials. In liquids, the velocity shows a pronounced resonance peak as a function of frequency. From the peak position and the use of the relationship $\omega_C \sim \eta^{-1}$, the static viscosity $\eta$ is inferred. In soft solids, the average rotation velocity is zero over the whole frequency range. The above predictions were tested using polysaccharide gellan gum aqueous dispersions (phytagel) at different weight concentrations between 0.2 and 1 wt. %. Without calcium chloride added, the polymers exhibit purely viscous or viscoelastic behaviors. With the addition of 1 mM of calcium chloride, the samples appear as gels above a critical concentration (around 0.2 wt. %). Polysaccharide gel samples were studied by cone-and-plate rheology and displays storage and loss moduli characteristic of soft solids, with storage moduli between 10 and 300 Pa. The stress *versus* rate curves exhibit a Herschel-Bulkley behavior, with yield stresses between 0.5 and 7 Pa. Both viscous and soft solid samples were studied using the MRS technique and the results confirm the predictions from the constitutive models. In particular for the yield stress materials, the wire average rotation velocity was found to be zero over broad frequency and elasticity ranges. From the oscillation amplitudes, the equilibrium elastic modulus of the gellan gum samples was estimated, and was found to obey the polymer dynamic theory.

The major differences between viscoelastic liquids and soft solids as emphasized by MRS lie in the experimental conditions. In particular, the values of the magnetic torques derived from Eq. 1 obviously play a central role. With wire lengths in the micron range, magnetic susceptibility around 1 and magnetic field of the order of 10 mT, the wires are subjected to torques of the order of $10^{-18}$ to $10^{-16}$ Nm. Such values are typically 8 orders of magnitude lower than the detection limit of most recent rheometers. These torques correspond to forces of the order of picoNewton and stresses of the order of milliPascal. As a result, the experiments performed with wires are in general associated with low deformations and low deformation rates. For Newton and Maxwell fluids, the flow properties are tested in the low shear rate range, leading to the static viscosity determination. For soft solids, the applied stress being lower than the yield stress, only the deformation regime is reached and the wire behavior reflects the strength of the gel elasticity. In conclusion, wire-based microrheology is a powerful technique able to determine the rheological nature of viscoelastic materials, and provide quantitative rheological parameters such as the static viscosity or the shear elastic modulus. Further developments are still needed to deal with more complex constitutive models such as the generalized Maxwell or Standard Linear Solid models to account for relaxation time dispersity present in real fluids. The technique could also start paving the way for a broad range of applications in rheology for samples available in small volumes or for samples that cannot be studied by rheometry.

# 5. Experimental Section

## 5.1 – Magnetic wires

Maghemite ($\gamma$-$Fe_2O_3$) nanoparticles were obtained by co-precipitation of iron(II) and iron(III) salts in aqueous solution follow by further oxidation of magnetite ($Fe_3O_4$).[61,62] The maghemite nanoparticle were characterized by transmission electron microscopy (Jeol-100 CX) and exhibit a mean diameter of 13.2 nm and a dispersity of 0.23, respectively (Supplementary Figures S3-S5).[63] The iron oxide nanoparticles were coated with poly(acrylic) acid polymers (Aldrich, $M_w$ = 5100 g $mol^{-1}$). Wires were synthesized by electrostatic co-assembly of negatively charged coated nanoparticles with poly(diallyldimethylammonium chloride) (Aldrich, $M_w$ > 100000 g $mol^{-1}$) polycations.[64] The bottom-up approach is based on the desalting transition of mixed solutions in the presence of a 0.3 T magnetic field, and using a dialysis cassette (Thermo Scientific, 10000 g $mol^{-1}$ membrane cutoff) at pH 8 (Supplementary Figure S6). Synthesized wires were autoclaved at 120°C and atmospheric pressure for 20 min to prevent bacterial contamination and stored at 4°C. Their length and diameter distributions were measured by optical microcopy (Olympus IX73) with a 100× objective lens and a CCD camera (QImaging, EXi Blue) working with Metaview (Universal Imaging). Both distributions were found to be well described by Log-Normal functions with median length 15.8 μm and median diameter 0.8 μm ($n$ = 86 and 102 respectively, see Supplementary Figure S7). The wires were also characterized with respect to their mechanical rigidity.[65] It was found that their persistence length was about 1 m, their Young modulus around 3 MPa, and that under rotation the wires do not exhibit detectable deformation or bending.[14,15]

## 5.2 – Polysaccharide gels

Phytagel (Sigma Alrich, P8169) is a linear polysaccharide composed of glucuronic acid, glucose and rhamnose monomers produced from *Pseudomonias elodea* bacterium.[57] Phytagel solutions (0.1 to 2.0 wt. %) were prepared by slow addition of the polysaccharide powder in 1 mM calcium chloride ($CaCl_2$, Fluka, purum) solutions under vigorous agitation.[58,59] The solutions were prepared with ultrapure deionized water (Millipore 18 MΩ cm, total organic content < 2 ppb) and heated at 70° C for 60 min to facilitate solubilization. The presence of divalent calcium ions also favors the gel formation, which arises from the complexation with carboxylates and hydroxyl functional groups.[58]. With increasing concentration and at the temperature of 30 °C, the biopolymer solutions exhibit a sol-gel transition around a critical concentration $c$ = 0.2 wt. % between a Newton fluid and an elastic gel (Figure 3a). Phytagel solution complex modulus $G^*(\omega)$ and stress *versus* shear rate curves were measured using an AntonPaar MCR302 and MCR500 rheometers equipped with a 1 mm Couette device (AntonPaar, Germany) for the liquid phase and a cone-and-plate (AntonPaar, CP50-1) device.

## 5.3 – Micro-rheology device and environment

For the sample preparation, 25 μL of phytagel solution containing wires at highly dilute concentration (1 pM) was deposited on a glass plate and sealed into to a Gene Frame® (Abgene/Advanced Biotech) dual adhesive system. The Gene Frame dimensions are 10×10×0.25 $mm^3$. At such concentrations, the distance between neighboring wires is large (> 50 μm) and the probes do not interact with each other, either hydrodynamically or magnetically. It was also checked that the probes studied were not close to the upper and lower glass slides of the measuring cell. Bright field time-lapse microscopy was used to monitor the wire actuation as a function of time. Stacks of images were acquired on the IX73 Olympus inverted microscope described previously. The glass plate was introduced into an home made device generating a magnetic rotational field, thanks to two pairs of coils (23 Ω) working with a 90°-phase shift (Supplementary Figure S8). An electronic set-up made of a frequency generator and of an amplifier allowed measurements in the range $10^{-2}$ - 100 rad $s^{-1}$ and at magnetic fields B = 0 – 15 mTesla. Images of wires were digitized and treated by the ImageJ software and plugins. For the wire magnetic property calibration, experiments were performed on a 86.6 wt. % water-glycerol mixture of static viscosity $\eta$ = 0.092 Pa s (T = 26 °C). For wires made from 13.2 nm particles and PADADMAC polymers, we found $\Delta\chi$ = 2.68 ± 0.27, and a magnetic susceptibility $\chi$ = 4.02 ± 0.40.

## Acknowledgments

We would like to thank L. Chevry, A. Cebers, M.-A. Fardin, A. Hallou and L. Vitorazi for fruitful discussions. The master students who participated to the research, L. Carvhalo, A. Conte-Daban, C. Leverrier, C. Lixi and N. K. Sampathkumar are also acknowledged. The Laboratoire de PHysico-chimie des Electrolytes et Nanosystèmes InterfaciauX (PHENIX, UMR Université Pierre et Marie Curie-CNRS n° 8234) is acknowledged for providing us with the magnetic nanoparticles. ANR (Agence Nationale de la Recherche) and CGI (Commissariat à l'Investissement d'Avenir) are gratefully acknowledged for their financial support of this work through Labex SEAM (Science and Engineering for Advanced Materials and devices) ANR 11 LABX 086, ANR 11 IDEX 05 02. This research was supported in part by the Agence Nationale de la Recherche under the contracts ANR-13-BS08-0015 (PANORAMA) and ANR-12-CHEX-0011 (PULMONANO).

**TOC image**

Magnetic Rotational Spectroscopy in a yield stress gel

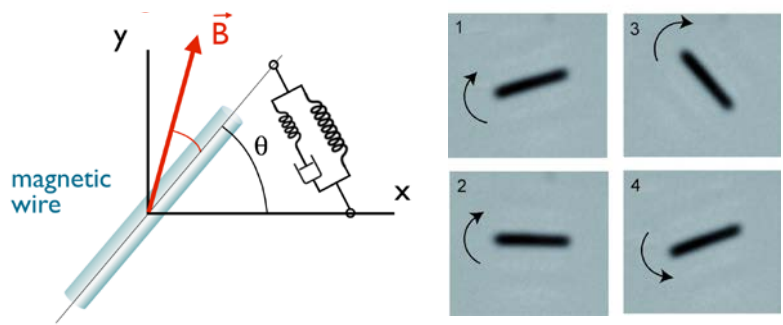